\documentclass[letterpaper, 10 pt, conference]{ieeeconf}  
\usepackage{graphicx}
\usepackage[hidelinks]{hyperref}
\usepackage{amsmath,amssymb,amsfonts}
\usepackage{booktabs}  
\usepackage{multirow}  
\usepackage{siunitx}
\usepackage{bm}
\usepackage[acronym]{glossaries}
\makeglossaries
    
\IEEEoverridecommandlockouts                              

\title{\LARGE \bf
Physics-Embedded  Neural Networks for sEMG-based \\Continuous Motion Estimation
}

\author{
	Wending Heng$^{1}$, 
	Chaoyuan Liang$^{1}$, 
	Yihui Zhao$^{2}$, 
	Zhiqiang Zhang$^{3}$, 
	Glen Cooper$^{1}$, 
	and Zhenhong Li$^{1,\dag}$ 
	\vspace{1em} \\ 
	$^{1}$University of Manchester\quad
	$^{2}$University of Bristol\quad
	$^{3}$University of Leeds
\thanks{$^{\dag}$ Corresponding Author (email:zhenhong.li@anchester.ac.uk)}
\thanks{IEEE/RSJ International Conference on Intelligent Robots and Systems}
}%

\newacronym{sEMG}{sEMG}{surface electromyography}
\newacronym{MSK}{MSK}{musculoskeletal}
 \newacronym{PINN}{PINN}{Physics-Informed Neural Network}
 \newacronym{PENN}{PENN}{Physics-Embedded Neural Network} 
  \newacronym{CNN}{CNN}{Convolutional Neural Network}
  \newacronym{LSTM}{LSTM}{Long Short-Term Memory}
  \newacronym{BPNN}{BPNN}{Backpropagation Neural Network}
  \newacronym{Bi-LSTM}{Bi-LSTM}{Bidirectional LSTM}
  \newacronym{MFTCAN}{MFTCAN}{Multi-Feature Temporal Convolutional Attention-Based Networks}
  \newacronym{MSE}{MSE}{mean square error}
  \newacronym{FEPI-PINN}{FEPI-PINN}{feature-encoded physics-informed neural network}
  \newacronym{RMSE}{$RMSE$}{root mean square error}
  \newacronym{RNN}{RNN}{Recurrent neural network}
  \newacronym{GAN}{GAN}{Generative Adversarial Network}
  
\begin{document}
\bstctlcite{IEEEexample:BSTcontrol}

\maketitle
\thispagestyle{empty}
\pagestyle{empty}

\begin{abstract}
Accurately decoding human motion intentions from surface electromyography (sEMG) is essential for myoelectric control and has wide applications in rehabilitation robotics and assistive technologies. However, existing sEMG-based motion estimation methods often rely on subject-specific musculoskeletal (MSK) models that are difficult to calibrate, or purely data-driven models that lack physiological consistency. This paper introduces a novel Physics-Embedded Neural Network (PENN) that combines interpretable  MSK forward-dynamics with data-driven residual learning, thereby preserving physiological consistency while achieving accurate motion estimation. The PENN employs a recursive temporal structure to propagate historical estimates and a lightweight convolutional neural network for residual correction, leading to robust and temporally coherent estimations. A two-phase training strategy is designed for PENN. Experimental evaluations on six healthy subjects show that PENN outperforms state-of-the-art baseline methods in both root mean square error (RMSE) and $R^2$ metrics. 
\end{abstract}

\section{Introduction}
\Gls{sEMG} has emerged as a powerful technique for decoding human motion intentions. Due to its noninvasive nature and relatively low cost, it holds significant potential for enabling intuitive and accessible human–machine interfaces in prosthetic device control \cite{jiangBioroboticsResearchNoninvasive2023}, rehabilitation robotics  \cite{teramaeEMGBasedModelPredictive2018d}, and human-robot collaboration \cite{xiongIntuitiveHumanRobotEnvironmentInteraction2024d, zengSimultaneouslyEncodingMovement2021}. However,  accurately  decoding continuous human motion in real time remains an ongoing challenge due to the complexities of   \gls{MSK} dynamics.

Existing model-free approaches use neural networks to directly estimate the mapping between \gls{sEMG} signals and joint kinematics \cite{baoCNNLSTMHybridModel2021}.  By neglecting underlying neuromechanical processes,  these ``black-box'' approaches may produce estimates that violate physical and \gls{MSK} constraints \cite{weiContinuousMotionIntention2024b,sitoleContinuousPredictionHuman2023b}, thereby raising reliability concerns in scenarios where physiological consistency is crucial.

Alternatively, model-based approaches establish an explicitly nonlinear transformation between  muscle activation, mechanical muscle forces, and joint movements by combining the muscle activation model,  the muscle-tendon dynamics model, and the joint dynamics model, preserving interpretability and ensuring physiological consistency \cite{copUnifyingSystemIdentification2021}.
Nevertheless, the model personalization requires a computationally intensive optimization process, which may span days due to high-dimensional search spaces and complex  \gls{MSK} dynamics \cite{sylvesterReviewMusculoskeletalModelling2021}. Simplifications such as rigid tendon assumptions are often adopted to reduce computational complexity; however, these simplifications compromise estimation accuracy \cite{fuArborSimArticulatedBranching2024}.

Recent hybrid approaches have demonstrated the potential of incorporating neuromechanical processes into model-free approaches for \gls{MSK} modeling.
Pioneering \glspl{PINN} have been proposed to incorporate joint dynamics as soft constraints within the loss function, allowing for the simultaneous estimation of muscle forces and joint angles \cite{zhangPhysicsInformedDeepLearning2023a}. Such integration guides the learning process towards physiologically consistent estimations. However, these approaches cannot guarantee the exact satisfaction of physiological constraints, especially when the physics penalty is not sufficiently large \cite{zouCorrectingModelMisspecification2024}. Moreover, the competition between physiological constraints and motion estimation can complicate the optimization process.

In this paper, we propose a new \gls{PENN}  for continuous motion estimation. It is composed of the physics-embedded module and the residual learning module, thereby effectively integrating \gls{MSK} forward dynamics with data-driven learning. Unlike conventional \glspl{PINN} with soft constraints, our approach incorporates a recursive structure to leverage temporal context from physical states and employs a streamlined \gls{CNN} to learn residual adjustments. The physics-embedded module facilitates backpropagation-driven optimization of muscle-tendon parameters, transforming the learning task from direct \gls{sEMG}-kinematics regression into a more tractable residual estimation problem. This strategy preserves physiological consistency while compensating for deficiencies in the \gls{MSK} model, thus achieving robust continuous motion decoding. Experimental validation on wrist flexion/extension tasks demonstrates that the proposed method outperforms state-of-the-art CNN-LSTM \cite{baoCNNLSTMHybridModel2021} and Bi-LSTM \cite{maBiDirectionalLSTMNetwork2021} models in terms of accuracy and temporal consistency, validating the efficacy of the proposed \gls{PENN} for \gls{sEMG}-driven continuous motion estimation.

\section{Related Work}
\label{sec:related_work}
This section reviews the related literature on  \gls{sEMG}-driven continuous motion estimation using model-free and hybrid approaches from the past five years.
\subsubsection{Model-free approaches} 
In \cite{maContinuousEstimationKnee2020},  a \gls{LSTM} network is introduced for knee joint estimation, effectively modeling temporal dependencies and yielding improved accuracy compared to traditional \gls{BPNN}. Based on this work, a \gls{Bi-LSTM} is proposed with enhanced temporal features \cite{maBiDirectionalLSTMNetwork2021}. However, while \gls{Bi-LSTM} excels in capturing temporal relationships, it fails to address spatial correlations among synergistic muscles, which are critical for multi-channel sEMG analysis. In fact, continuous motion estimation is inherently spatiotemporal, necessitating the capture of temporal muscle activation patterns and spatial dependencies between muscles.
To address this limitation, a CNN-LSTM model is introduced by combining  \gls{CNN} for spatial feature extraction with \gls{LSTM} for temporal pattern modeling \cite{baoCNNLSTMHybridModel2021}. This approach significantly outperformed standalone \gls{CNN} and \gls{LSTM} in wrist motion estimation. Subsequent advancements have been made with the \gls{MFTCAN}, which integrates temporal convolutions and attention mechanisms to learn  temporal dependencies from \gls{sEMG} and feature contributions at different moments \cite{wangContinuousEstimationHuman2022}. 


\begin{figure*}[!t]
  \centering
  \includegraphics[width=1\textwidth]{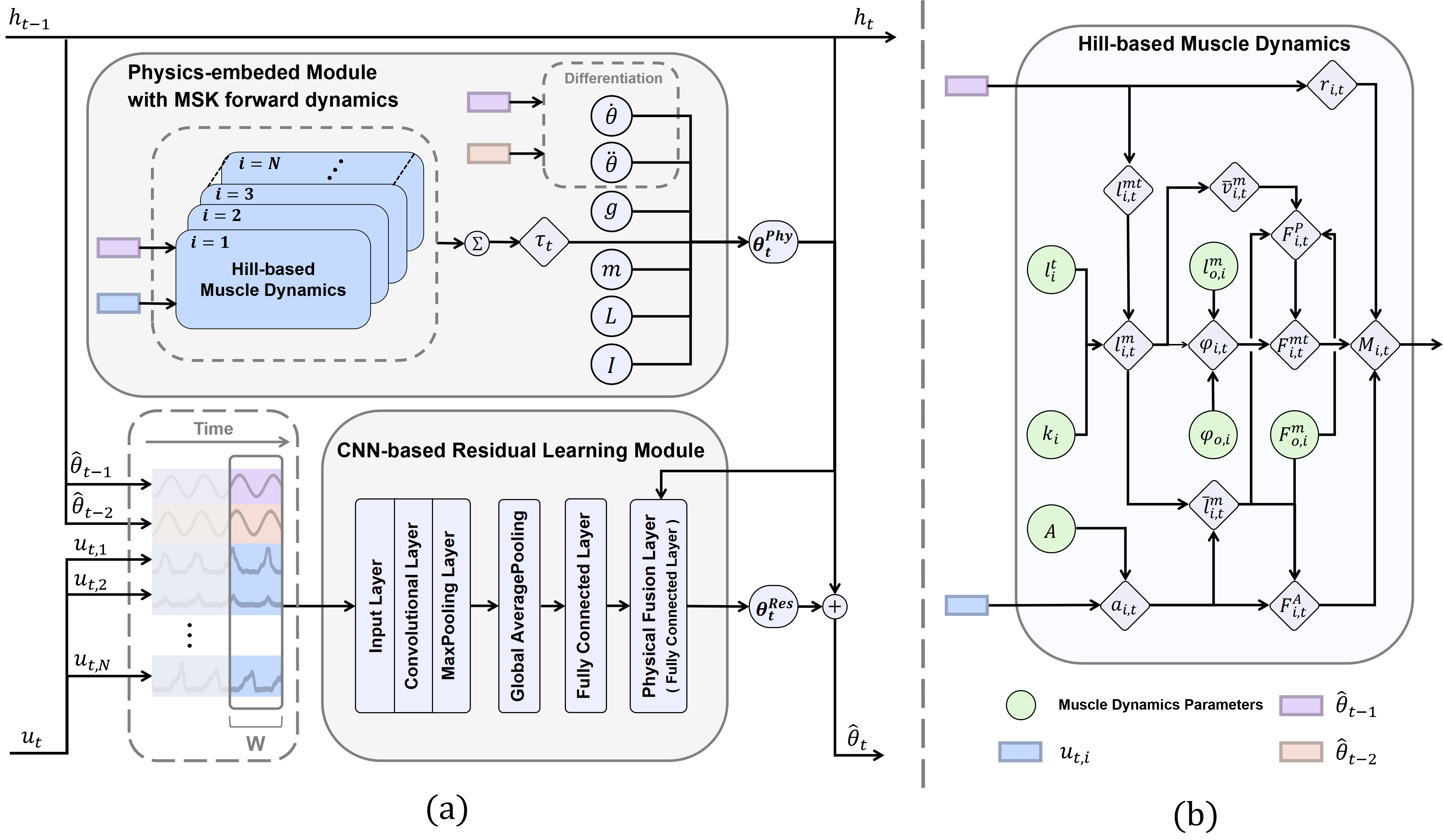}
  \caption{
  $(a)$ \gls{PENN} framework. It is composed of a Physics-embedded Module and a Residual Learning Module. It uses a recursive structure for the explicit propagation of temporal information. \gls{PENN} uses historical estimates $h_{t-1} = \{ \hat\theta_{t-1}, \hat\theta_{t-2}\}$ and pre-processed \gls{sEMG} signal $u_t = [u_{t,i}]_{i=1}^N$ to generate the estimate $\hat\theta_t$. 
  $(b)$ Hill-based muscle dynamics of $i$th muscle.
  }
  \label{fig_framework}
\end{figure*}

\subsubsection{Hybrid approaches}
 To mitigate errors from inaccurate joint dynamics parameters, a novel \gls{PINN} is proposed in \cite{zhangPhysicsInformedDeepLearning2023a}  by incorporating a joint dynamics model into the traditional angle-based loss function. In \cite{shiPhysicsInformedLowShotAdversarial2024a}, the joint dynamics model is integrated with the adversarial learning framework as a ``physiological consistency reward''. Both approaches enhance estimation accuracy by infusing physical information into the network. Since  physiological inconsistency stems from factors beyond joint dynamics alone, \gls{FEPI-PINN} in \cite{tanejaFeatureEncodedPhysicsInformedParameter2022a} extends PINNs by introducing \gls{MSK} forward dynamics into the loss function as soft constraints. This hybrid formulation, which combines estimation error with \gls{MSK} constraints, guides networks to reduce physiological inconsistency. The \gls{FEPI-PINN} is further enhanced by introducing wavelet decomposition to handle high-frequency noise in \gls{sEMG} signals. Based on hybrid loss functions, 
 several \glspl{PINN} have been developed \cite{zhangBoostingPersonalizedMusculoskeletal2023, maPhysicsInformedDeepLearning2024a}, and have achieved improved accuracy and physiological consistency. However, soft constraints in the loss function cannot guarantee the exact satisfaction of physiological constraints. The competition between \gls{MSK} constraints and estimation loss complicates parameter tuning and optimization.

\section{Methods}
Consider the one-degree-of-freedom joint rotational motion driven by $N$ muscle-tendon units. The objective of this paper is to estimate the joint rotation angle $\theta$ using \gls{sEMG} signals.
\subsection{\gls{PENN} Framework }
The \gls{PENN} framework proposed in this paper adopts a hybrid architecture, comprising a physics-embedded module and a \gls{CNN}-based residual learning module, as shown in Fig.~\ref{fig_framework}. In the physics-embedded module, \gls{MSK} forward dynamics are integrated to provide physiologically inconsistent motion estimation. The \gls{CNN}-based residual learning module captures nonlinear residuals that the physics-embedded module cannot model, effectively bridging the gap between \gls{MSK} forward dynamics and the complex EMG-to-kinematics mapping.  It adopts a recursive temporal context integration strategy, where pre-processed sEMG signals at the current time step and historical estimations of joint angles at the previous two time steps are input into the recursive structure unit. Then, the current output of the physics-embedded module is incorporated into the physical fusion layer (a fully connected layer) to enhance residual learning.
  A unified total loss function is constructed based on the physics-embedded loss and the residual learning loss using \gls{MSE}.

\subsection{\gls{PENN} Architecture}
 The \gls{PENN} architecture is composed of two following modules as shown in Fig.~\ref{fig_framework}.
\subsubsection{Physics-Embedded Module}
This module leverages a Hill-based forward dynamics model to generate an intermediate estimate  $\theta^{Phy}_{t}$ using previous estimates $\hat{\theta}_{t-1}$, $\hat{\theta}_{t-2}$ and pre-processed \gls{sEMG} signals $u_{i,t}$, where $t$ is the time step and $i=1,\cdots,N$  is the index of the muscle. The Hill-based forward dynamics model in this paper includes Hill-type muscle models (muscle activation models, muscle-tendon dynamics models) and a joint dynamics model. Hill-type muscle models are widely used in model-based approaches to describe the \gls{sEMG}-force relationship for individual muscles \cite{zhaoEMGDrivenMusculoskeletalModel2020}. To reduce numerical stiffness in the muscle-tendon dynamics model, we assume the tendon to be rigid, which implies that the pennated muscle element, comprising a contractile element in parallel with a passive elastic element, is connected to an inextensible tendon element \cite{millard2013flexing}.  The following provides a detailed explanation of the Hill-based forward dynamics model.

\noindent\textbf{\textit{Muscle Activation Model:}}
For the $i$th muscle,  the nonlinear relationship between the pre-processed sEMG signal $u_{i,t}$ and muscle activation  $a_{i,t}$ is given as 
\begin{align}
    a_{i,t} = \frac{e^{A u_{i,t}} - 1}{e^A - 1} \quad 
\end{align}
where $A$ represents the coefficient to account for nonlinearity.

\noindent \textbf {\textit{Muscle-Tendon Dynamics model:}} 
The muscle-tendon force $F^{mt}_{i,t}$ is given by
\begin{align}
F^{mt}_{i,t} = \left(F^a_{i,t}+F^p_{i,t}\right)\cos\varphi_{i,t}
\end{align}
where $F^a_{i,t}$ is the active muscle contraction force and $F^p_{i,t}$ is the passive force generated by muscle stretch, and $\varphi_{i,t}$ is the pennation angle between the muscle fiber
and tendon. The pennation angle is calculated by
\begin{align}
\varphi_{i,t}=\arcsin(\frac{l_{o,i}^m\sin\varphi_{o,i}}{l^m_{i,t}})
\end{align}
where $l^m_{i,t}$ is the muscle fiber length, $l_{o,i}^m$ is the optimal muscle fiber length and $\varphi_{o,i}$ is the optimal pennation angle.
 
The active muscle contraction force $F^a_{i,t}$ can be calculated as
\begin{align}
F^a_{i,t}=F_{o,i}^mf^a\left(\bar{l}_{i,t}^a\right)f\left(\overline{v}_{i,t}\right)a_{i,t}
\end{align}
where $F^m_{o,i}$ is the maximum isometric force, $f^a\left(\bar{l}_{i,t}^a\right)$ is the active force-length characteristics,  $f(\bar{v}_{i,t})$ is the  force-velocity characteristics of the muscle, and 
\begin{align}
    \bar{l}_{i,t}^a = \frac{l^t_{i,t}}{l^m_{o,i}(\lambda(1-a_{i,t})+1)}
\end{align}
 is the normalized muscle fiber length relative to the corresponding activation. $\lambda$ is a constant and is set to 0.15 according to \cite{lloydEMGdrivenMusculoskeletalModel2003}.

The active force-length characteristics is given as 
\begin{align}
f_a(\bar{l}_{i,t}^a)=e^{-(\bar{l}_{i,t}^a-1)^2k^{-1}}
\end{align}
where $k$ is a constant and is set to 0.45 according to \cite{thelenAdjustmentMuscleMechanics2003}. 

The force-velocity characteristics is given as 
\begin{align}
f(\overline{v}_{i,t})=\left\{
\begin{array}
{ll}\frac{0.3(\overline{v}_{i,t}+1)}{-\overline{v}_{i,t}+0.3} & \overline{v}_{i,t}\leq0 \\
\frac{2.34\overline{v}_{i,t}+0.039}{1.3\overline{v}_{i,t}+0.039} & \overline{v}_{i,t}>0
\end{array}\right.
\end{align}
where $\bar{v}_{i,t} = v_{i,t}/v_{o,i}$ is the normalized muscle contraction velocity, $v_{i,t}$ is the derivative of the muscle fiber length  and $v_{o,i}$ is the maximum contraction velocity. $v_{o,i}$ is set to $10  l^m_{i,t}/sec$ according to \cite{zajacMuscleTendonProperties}.

The passive muscle stretch force is given as
\begin{align}
F^p_{i,t}=\left\{
\begin{array}
{ll}0 & l_{i,t}^m\leq l_{o,i}^m \\
f_{p}\left(\bar{l}_{i}^{m}\right) F_{o, i}^{m}& l_{i,t}^m>l_{o,i}^m
\end{array}\right.
\end{align}
where $\bar{l}_{i,t}^m = l^m_{i,t} / l^m_{o,i}$ is the normalized muscle fiber length,  and 
\begin{align}
    f_{p}\left(\bar{l}_{i}^{m}\right)=\frac{e^{10\left(\bar{l}_{i}^{m}-1\right)}}{e^{5}}
\end{align}
 is passive force-length characteristics. 
 
\noindent \textbf {\textit{ Joint Dynamics Model:}}
For single degree-of-freedom joint motions, we assume the joint to be a single hinge joint. The dynamics of the joint can be described as

\begin{equation}
\tau = I\ddot{\theta}+C\dot{\theta}+mgL\sin\theta
\end{equation}
where $I$ represents the moment of inertia,  $C$ is the damping coefficient accounting for the elastic and viscous effects of the tendons and ligaments, $m$ is the mass, $L$ is the length between the rotation center and the center of mass, and $\tau$ is the total joint torque.   $\tau$ can also be calculated as 

\begin{equation}
\tau_t=\sum_{n=1}^Nr_{i,t}F^{mt}_{i,t}
\end{equation}
where $r_{i,t}$ is the moment arm of $i$th muscle which is 
calculated using polynomials and the scale coefficient based on OpenSim \cite{sethOpenSimSimulatingMusculoskeletal2018}.

\subsubsection{\gls{CNN}-based Residual Learning Module}
 The proposed residual learning \gls{CNN} employs a lightweight architecture composed of a convolutional block, a max pooling layer, a fully connected layer, and a physical fusion layer.
 The CNN processes a sliding window of $W$ time steps, comprising  $N+2$ features ($N$ pre-processed \gls{sEMG} signals $u_{i,t}$ and two historical estimates, $\hat{\theta}_{t-1}$ and $\hat {\theta}_{t-2}$). These features collectively form an $(N+2) \times W$ input matrix.
 In the convolutional block, a one-dimensional convolution layer ($N+2$ input channels, 32 output channels, kernel size 3, stride 1, padding 1) is used to preserve temporal resolution, followed by ReLU activation.  A max pooling layer (kernel size 2, stride 1, padding 1) reduces dimensionality while retaining key temporal features. Dropout is used to further mitigate overfitting, and a global average pooling layer condenses the feature map into a fixed-length vector. This vector is then passed through a fully connected layer with 32 hidden nodes, followed by ReLU and dropout. The resulting output is concatenated with $\theta_t^{Phy}$ and fed to the physical fusion layer (which is a fully connected layer) to generate $\theta_t^{Res}$.

\subsection{Loss Function Design}
In this study, a unified loss function is designed to integrate MSK forward dynamics with residual learning. The total loss consists of two complementary components
\begin{equation}
L_{total} = \lambda L_{phy}+\beta L_{res}
\label{eqnLabel}
\end{equation}
where $\lambda$ and $\beta$ are the weights for the physical loss $L_{phy}$ and the residual loss $L_{res}$, respectively.  $L_{phy}$ minimizes the discrepancy between the intermediate estimate from physics-embedded module and the true values, and is defined as
\begin{equation}
L_{phy} = \frac{1}{T} \sum_{t=1}^{T} \left( \theta^{Phy}_{t}- \theta_{t} \right)^2
\end{equation}
where $\theta_t$ denotes the ground truth at time $t$. $L_{res}$ refines model performance by learning the residual errors and is defined as
\begin{equation}
L_{res} = \frac{1}{T} \sum_{t=1}^{T} \left( \theta^{Res}_{t} + \theta^{Phy}_{t} - \theta_{t} \right)^2
\end{equation}
In the following section, we will show how to train \gls{PENN} use $L_{total}$.

\subsection{\gls{PENN} Training}
We propose a two-phase training strategy for \gls{PENN}. Phase one focuses on the optimization of muscle dynamics parameters in the physics-embedded module ($F_{o,i}^m$, $l_i^t$, $k_i$, $l^m_{o,i}$, $\varphi_{o,i}$, and $A$). In this phase, the parameters of the physical fusion layer are initialized as zeros, and the weights in the loss function \eqref{eqnLabel} are set to  $\lambda=1$ and $\beta=0$. The training batch size is set to 1. Once the loss of the physics-embedded model converges,  phase two training starts, and the weights of the loss function \eqref{eqnLabel} are set to $\lambda=0$ and $\beta=1$. During this phase, the residual module learns the discrepancies between the intermediate estimate from the physics-embedded module $\theta^{Phy}_{t}$ and the ground truth $\theta_t$. The training batch size is set to $32$. For both phases, the training is conducted using the Adam optimizer with an initial learning rate of 0.001. The dropout rate is set to 0.3 and the window length $W$ is set to 16. During training, the dataset is split into training and test sets with a ratio of 8.5:1.5 , and a early stopping mechanism is implemented, terminating the training if no improvement is observed for 30 consecutive epochs.

\section{Datasets and Experimental Setting}
This section introduces data collection and pre-processing process, evaluation criteria and baseline methods used for evaluation. 

\subsection{Data Collection and Pre-processing}
Motion estimation of the wrist joint during flexion/extension tasks is used to evaluate the proposed \gls{PENN}. As approved by the MaPS and Engineering Joint Faculty Research Ethics Committee of the University of Leeds (MEEC 18-002), six subjects are recruited, and all participants have signed consent forms. During the experiment, participants are informed to perform wrist flexion/extension motions in a seated position, with 16 reflective markers placed on the right arm, as shown in Fig.~\ref{fig:ExpAFeriment}. Each subject performs five repetitive trials, with a three-minute rest between trials to prevent muscle fatigue. Marker trajectories are recorded (VICON motion capture system, 250 Hz) and filtered (Butterworth second, 1 Hz) to compute wrist kinematics. 
\begin{figure}[!t]
    \centering
\includegraphics[width=0.48\textwidth]{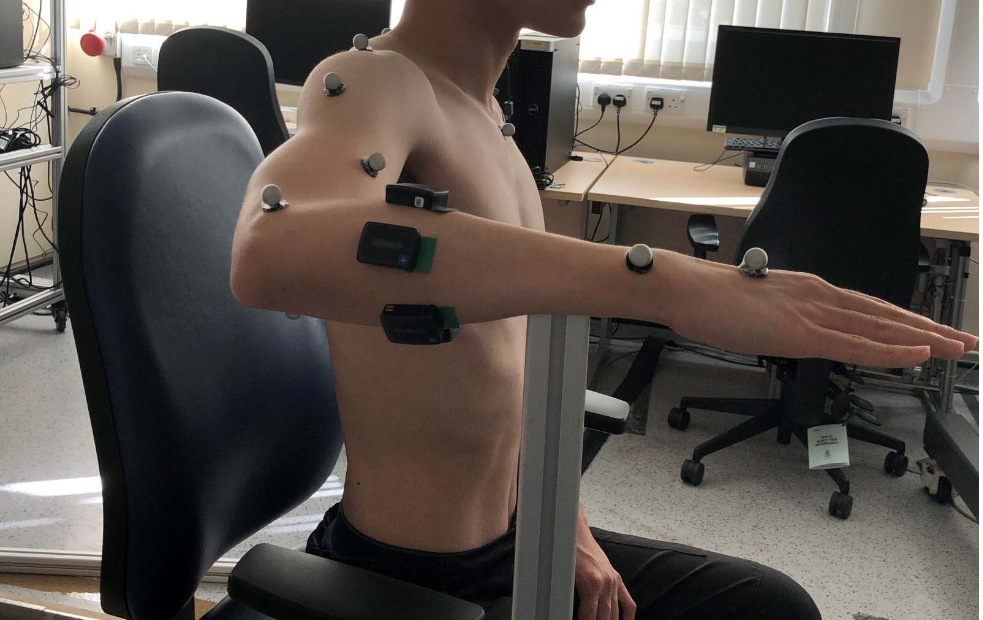}
    \vspace{-1em}
    \caption{Experiment setup: 16 reflective markers are attached on subject’s right upper limb. Electrodes are placed on five primary muscles of wrist joint, including FCR, FCU, ECU, ECRL and ECRB..}
    \label{fig:Experiment}
\end{figure}

sEMG signals of five wrist muscles are recorded (Delsys Avanti Sensors, 2000 Hz): Flexor Carpi Radialis (FCR), Flexor Carpi Ulnaris (FCU), Extensor Carpi Radialis Longus (ECRL), Extensor Carpi Radialis Brevis (ECRB), and Extensor Carpi Ulnaris (ECU). Raw EMG signals are bandpass-filtered (Butterworth fourth-order, 20–450 Hz), fully rectified, and low-pass-filtered (Butterworth fourth-order, 4 Hz). The resultant signals are normalized with respect to maximum voluntary contraction (MVC) that are measured
prior to the experiment. All the data are synchronized and resampled at 1000 Hz.

\subsection{Evaluation Criteria}
 \Gls{RMSE} and the coefficient of determination $(R^2)$ are used to evaluate the amplitude discrepancy and correlation between the estimated value and the ground truth, respectively. 
\begin{align}
RMSE&=\sqrt{\frac{1}{T}\sum_{t=1}^T(\theta_t-\hat{\theta}_t)^2} \\
R^2&=1-\frac{\sum_{t=1}^T(\theta_t-\hat{\theta}_t)^2}{\sum_{t=1}^T(\theta_t-\bar{\theta}_t)^2}
\end{align}
where $T$ is the number of samples.

\subsection{Baseline Methods}
To evaluate the effectiveness of \gls{PENN}, CNN-LSTM and Bi-LSTM models are selected as baseline models for comparison. Both models use \gls{MSE} loss function and the Adam optimizer for training, with an initial learning rate of $0.001$ and a dropout rate of $0.3$.
\subsubsection{CNN-LSTM} 
This method combines a \gls{CNN} encoder with a \gls{LSTM} decoder for spatiotemporal modeling. The \gls{CNN} encoder consists of four convolutional layers with ReLU activations, where the number of filters increases from 16 to 32 to capture spatial patterns from sEMG channels. Global average pooling is applied to reduce each feature map to a single neuron, followed by a fully connected layer that projects the features into latent space. These features are then repeated along the temporal axis and fed into the \gls{LSTM} decoder, which comprises two layers with 50 hidden neurons each. The sequence length is set to 200 for temporal processing, and the batch size is set to 64. 
\subsubsection{Bi-LSTM} 
This method captures temporal dependencies in \gls{sEMG} signals using a two-layer \gls{Bi-LSTM} architecture. Each \gls{Bi-LSTM} layer has 64 hidden units per direction, resulting in 128 hidden neurons per layer. The output from the last time step in both forward and backward directions is concatenated, forming a 128-dimensional feature vector. This feature vector is sent to a fully connected layer with one output neuron, representing the estimated wrist angle. A window size of 32 is used for temporal sequence modeling, allowing the model to capture temporal dependencies over 32 time steps, and the batch size is set to 32.

\section{Results}
\gls{PENN}, CNN-LSTM and Bi-LSTM are implemented on a workstation with a GeForce RTX 4070 Ti graphics card and 32 GB RAM.

\subsection{Convergence of Phased Learning}
\begin{figure}[!t]
    \centering
\includegraphics[width=0.48\textwidth]{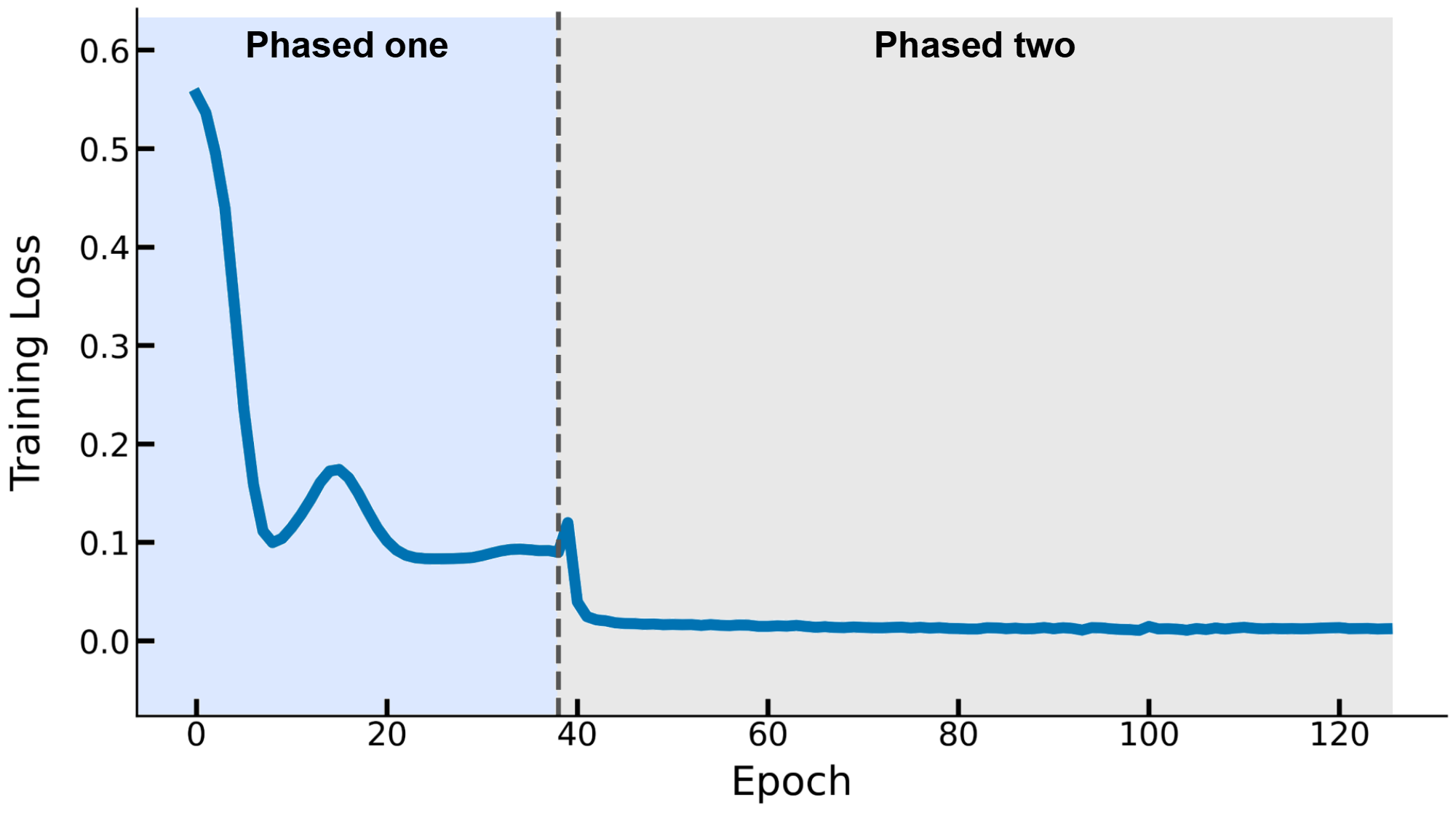}
    \vspace{-1em}
    \caption{Value of loss function $L_{total}$ during training.}
    \label{fig:loss}
\end{figure}

Fig.~\ref{fig:loss} presents the loss evolution during two-phase learning for the third subject.
In Phase One, the training loss decreases rapidly and exhibits fluctuations between 10-20 epochs due to the exclusive optimization of the Hill-muscle parameters.  These fluctuations may arise from the nonlinearity in the Hill-muscle model and the coupling parameters among multiple muscles. After convergence is reached, the model transitions to Phase Two at epoch 38 for residual learning. In this phase, the training loss decreases smoothly and stabilizes, demonstrating the stability of the two-phase learning strategy.

\subsection{Performance Comparison}
\begin{figure}[!t]
    \centering
\includegraphics[width=0.48\textwidth]{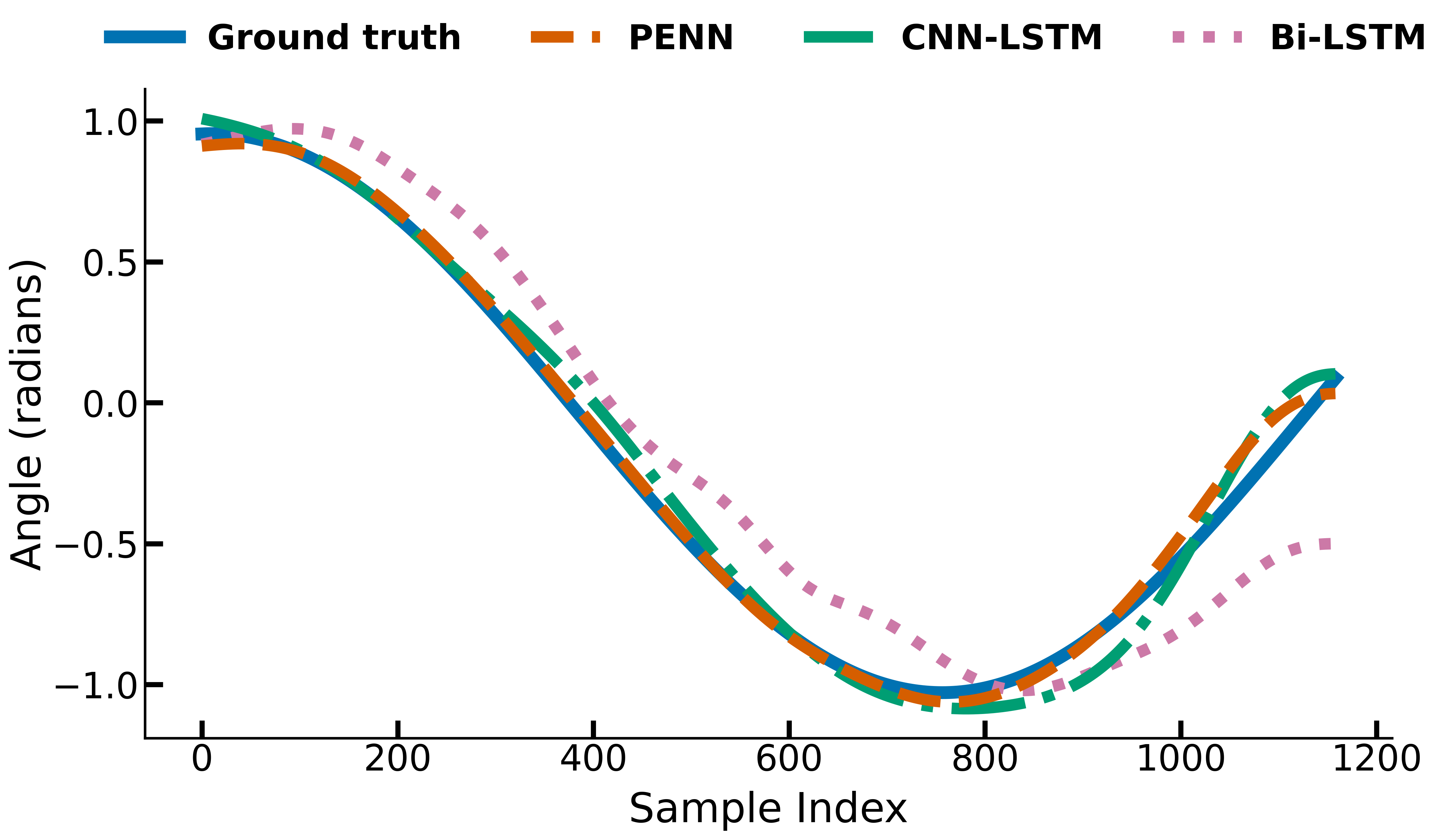}
    \vspace{-1em}
    \caption{Wrist joint angle estimated by PENN, CNN-LSTM, and Bi-LSTM for Subject two.}
    \label{fig:comparison}
\end{figure}

 Fig.~\ref{fig:comparison} shows the ground truth wrist angle $\theta$ and estimated angle $\hat{\theta}$ from \gls{PENN} and the baseline methods. The angle trajectory estimated by \gls{PENN}  closely aligns with the ground truth, maintaining temporal coherence and minimizing phase lag.  In contrast, the baseline models show larger deviations and inconsistencies, particularly during dynamic transitions. These results demonstrate the improved estimation accuracy and temporal consistency of \gls{PENN}. 

\begin{figure}[!t]
    \centering
    \includegraphics[width=0.48\textwidth]{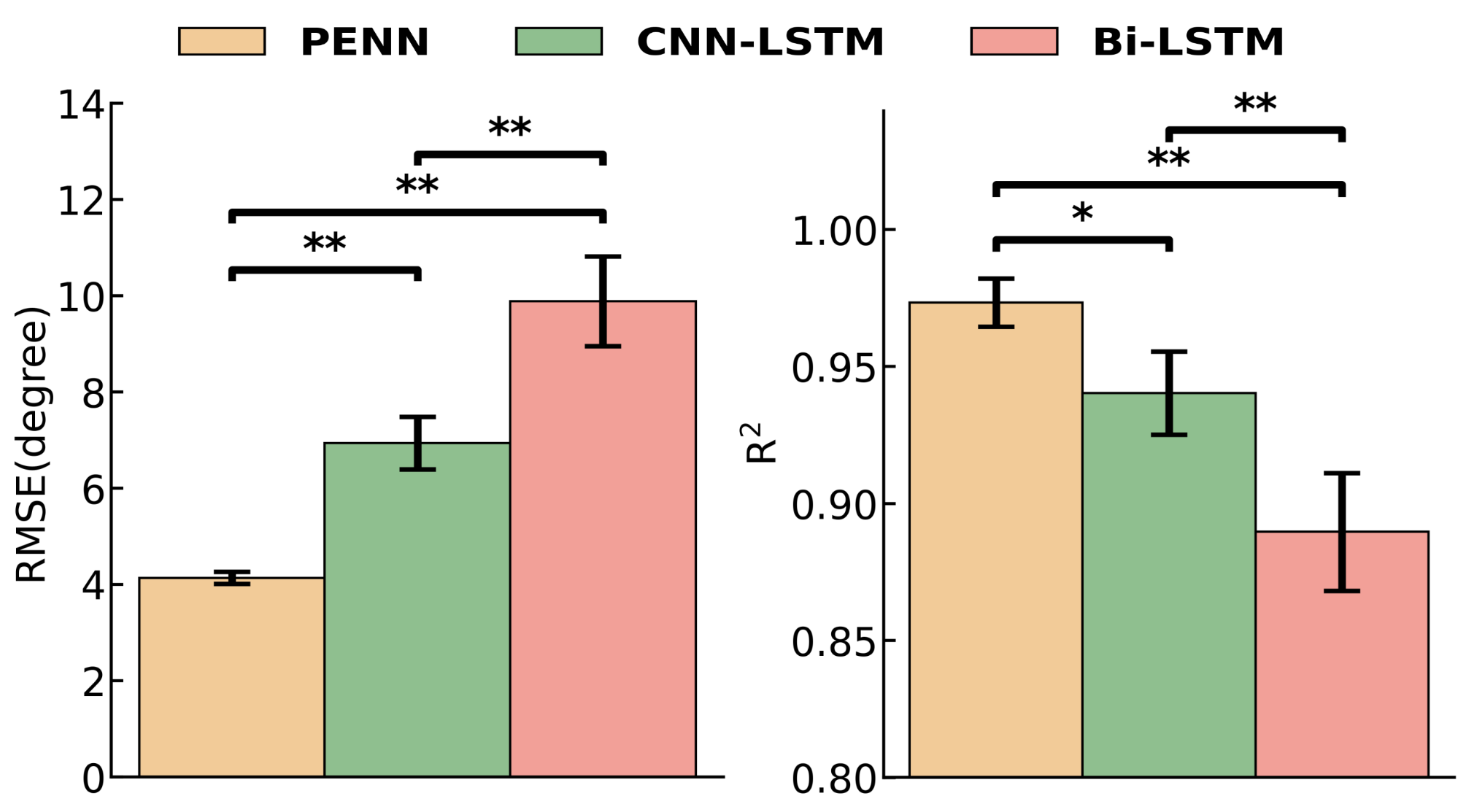} 
    \caption{Average \gls{RMSE} and $R^2$ of \gls{PENN}, CNN-LSTM, and Bi-LSTM.   Asterisks indicate the statistically significant difference between two methods (* for $p < 0.05$, ** for $p < 0.01$).}
    \label{fig:Paired_t}
\end{figure}

\Gls{RMSE} and  $R^2$ of \gls{PENN} and baseline methods are summarized in Table~\ref{tab:rmse_r2_comparison}. It shows that
\gls{PENN} consistently achieves the lowest \Gls{RMSE} and the highest $R^2$ across six subjects. To show the differences among three methods, paired t-tests are conducted and results are shown in Fig.~\ref{fig:Paired_t}.
It shows that \gls{PENN} not only significantly improves estimation accuracy, but also exhibits consistently lower variance across subjects, which implies enhanced robustness to inter-subject differences.


\begin{table}[!t]
    \centering
    \caption{ RMSE and $R^2$ of \gls{PENN} and Baseline Methods for Six Subjects}
    \label{tab:rmse_r2_comparison}
    \resizebox{0.48\textwidth}{!}{
    \begin{tabular}{ccccccc}
    \toprule
\textbf{Subject} & \multicolumn{2}{c}{\textbf{PENN}} & \multicolumn{2}{c}{\textbf{CNN-LSTM}} & \multicolumn{2}{c}{\textbf{Bi-LSTM}} \\
                 & \textbf{RMSE} & $\bm{R^2}$ & \textbf{RMSE} & $\bm{R^2}$ & \textbf{RMSE} & $\bm{R^2}$ \\
\midrule
\textbf{S1} & \textbf{3.79} & \textbf{0.993} & 7.40 & 0.976 & 11.87 &  0.924\\
\textbf{S2} & \textbf{4.50} & \textbf{0.984} &  6.29 & 0.980 & 7.14 & 0.974 \\
\textbf{S3} & \textbf{4.14} & \textbf{0.987} & 7.79 & 0.953 & 12.27 & 0.883 \\
\textbf{S4} & \textbf{3.84} & \textbf{0.933} & 4.83 & 0.893 & 7.14 & 0.841 \\
\textbf{S5} & \textbf{4.47} & \textbf{0.971} & 6.62 & 0.900 & 11.01 & 0.834 \\
\textbf{S6} & \textbf{4.07} & \textbf{0.972} & 8.68 & 0.940 & 9.87 & 0.882 \\
\midrule
\textbf{Average}   & \textbf{4.13} & \textbf{0.973} & 6.94 & 0.940 & 9.88 & 0.890 \\
\bottomrule
    \end{tabular}
    }
\end{table}

Embedding muscle forward dynamics with neural network, enhances the interpretability of the network and reduces the complexity of the mapping that needs to be modeled by the residual learning module, thus resulting in more accurate and temporally coherent estimations.

\section{CONCLUSIONS}
This paper proposes a novel \gls{PENN} for sEMG-driven continuous motion estimation. It employs a recursive structure and combines MSK forward dynamics with data-driven residual learning, enhancing accuracy while preserving physiological consistency. A two-phase training strategy is designed for \gls{PENN}. The experimental results demonstrate the effectiveness of \gls{PENN}, as evidenced by its superior performance in wrist flexion/extension tasks compared to state-of-the-art CNN-LSTM and Bi-LSTM approaches.

\addtolength{\textheight}{-12cm}   




\bibliographystyle{IEEEtran}
\bibliography{IROS}

\begin{thebibliography}{10}
\providecommand{\url}[1]{#1}
\csname url@samestyle\endcsname
\providecommand{\newblock}{\relax}
\providecommand{\bibinfo}[2]{#2}
\providecommand{\BIBentrySTDinterwordspacing}{\spaceskip=0pt\relax}
\providecommand{\BIBentryALTinterwordstretchfactor}{4}
\providecommand{\BIBentryALTinterwordspacing}{\spaceskip=\fontdimen2\font plus
\BIBentryALTinterwordstretchfactor\fontdimen3\font minus
  \fontdimen4\font\relax}
\providecommand{\BIBforeignlanguage}[2]{{%
\expandafter\ifx\csname l@#1\endcsname\relax
\typeout{** WARNING: IEEEtran.bst: No hyphenation pattern has been}%
\typeout{** loaded for the language `#1'. Using the pattern for}%
\typeout{** the default language instead.}%
\else
\language=\csname l@#1\endcsname
\fi
#2}}
\providecommand{\BIBdecl}{\relax}
\BIBdecl

\bibitem{jiangBioroboticsResearchNoninvasive2023}
N.~Jiang, C.~Chen, J.~He, J.~Meng, L.~Pan \emph{et~al.}, ``Bio-robotics
  research for non-invasive myoelectric neural interfaces for upper-limb
  prosthetic control: A 10-year perspective review,'' \emph{National Science
  Review}, vol.~10, no.~5, p. nwad048, Apr. 2023.

\bibitem{teramaeEMGBasedModelPredictive2018d}
T.~Teramae, T.~Noda, and J.~Morimoto, ``{{EMG-Based Model Predictive Control}}
  for {{Physical Human}}--{{Robot Interaction}}: {{Application}} for
  {{Assist-As-Needed Control}},'' \emph{IEEE Robotics and Automation Letters},
  vol.~3, no.~1, pp. 210--217, Jan. 2018.

\bibitem{xiongIntuitiveHumanRobotEnvironmentInteraction2024d}
D.~Xiong, D.~Zhang, Y.~Chu, Y.~Zhao, and X.~Zhao, ``Intuitive
  {{Human-Robot-Environment Interaction}} with {{EMG Signals}}: {{A Review}},''
  \emph{IEEE/CAA Journal of Automatica Sinica}, vol.~11, no.~5, pp. 1075--1091,
  May 2024.

\bibitem{zengSimultaneouslyEncodingMovement2021}
C.~Zeng, C.~Yang, H.~Cheng, Y.~Li, and S.-L. Dai, ``Simultaneously {{Encoding
  Movement}} and {{sEMG-Based Stiffness}} for {{Robotic Skill Learning}},''
  \emph{IEEE Transactions on Industrial Informatics}, vol.~17, no.~2, pp.
  1244--1252, Feb. 2021.

\bibitem{baoCNNLSTMHybridModel2021}
T.~Bao, S.~A.~R. Zaidi, S.~Xie, P.~Yang, and Z.-Q. Zhang, ``A {{CNN-LSTM Hybrid
  Model}} for {{Wrist Kinematics Estimation Using Surface Electromyography}},''
  \emph{IEEE Transactions on Instrumentation and Measurement}, vol.~70, pp.
  1--9, 2021.

\bibitem{weiContinuousMotionIntention2024b}
Z.~Wei, Z.-Q. Zhang, and S.~Q. Xie, ``Continuous {{Motion Intention Prediction
  Using sEMG}} for {{Upper-Limb Rehabilitation}}: {{A Systematic Review}} of
  {{Model-Based}} and {{Model-Free Approaches}},'' \emph{IEEE Transactions on
  Neural Systems and Rehabilitation Engineering}, vol.~32, pp. 1487--1504,
  2024.

\bibitem{sitoleContinuousPredictionHuman2023b}
S.~P. Sitole and F.~C. Sup, ``Continuous {{Prediction}} of {{Human Joint
  Mechanics Using EMG Signals}}: {{A Review}} of {{Model-Based}} and
  {{Model-Free Approaches}},'' \emph{IEEE Transactions on Medical Robotics and
  Bionics}, vol.~5, no.~3, pp. 528--546, Aug. 2023.

\bibitem{copUnifyingSystemIdentification2021}
C.~P. Cop, G.~Cavallo, R.~C. Van 'T~Veld, B.~Fjm~Koopman, J.~Lataire
  \emph{et~al.}, ``Unifying system identification and biomechanical
  formulations for the estimation of muscle, tendon and joint stiffness during
  human movement,'' \emph{Progress in Biomedical Engineering}, vol.~3, no.~3,
  p. 033002, Jul. 2021.

\bibitem{sylvesterReviewMusculoskeletalModelling2021}
A.~D. Sylvester, S.~G. Lautzenheiser, and P.~A. Kramer, ``A review of
  musculoskeletal modelling of human locomotion,'' \emph{Interface Focus},
  vol.~11, no.~5, p. 20200060, Oct. 2021.

\bibitem{fuArborSimArticulatedBranching2024}
X.~Fu, J.~Withers, J.~A. Miyamae, and T.~Y. Moore, ``{{ArborSim}}:
  {{Articulated}}, branching, {{OpenSim}} routing for constructing models of
  multi-jointed appendages with complex muscle-tendon architecture,''
  \emph{PLOS Computational Biology}, vol.~20, no.~7, p. e1012243, Jul. 2024.

\bibitem{zhangPhysicsInformedDeepLearning2023a}
J.~Zhang, Y.~Zhao, F.~Shone, Z.~Li, A.~F. Frangi \emph{et~al.},
  ``Physics-{{Informed Deep Learning}} for {{Musculoskeletal Modeling}}:
  {{Predicting Muscle Forces}} and {{Joint Kinematics From Surface EMG}},''
  \emph{IEEE Transactions on Neural Systems and Rehabilitation Engineering},
  vol.~31, pp. 484--493, 2023.

\bibitem{zouCorrectingModelMisspecification2024}
Z.~Zou, X.~Meng, and G.~E. Karniadakis, ``Correcting model misspecification in
  physics-informed neural networks ({{PINNs}}),'' \emph{Journal of
  Computational Physics}, vol. 505, p. 112918, May 2024.

\bibitem{maBiDirectionalLSTMNetwork2021}
C.~Ma, C.~Lin, O.~W. Samuel, W.~Guo, H.~Zhang \emph{et~al.}, ``A
  {{Bi-Directional LSTM Network}} for {{Estimating Continuous Upper Limb
  Movement From Surface Electromyography}},'' \emph{IEEE Robotics and
  Automation Letters}, vol.~6, no.~4, pp. 7217--7224, Oct. 2021.

\bibitem{maContinuousEstimationKnee2020}
X.~Ma, Y.~Liu, Q.~Song, and C.~Wang, ``Continuous {{Estimation}} of {{Knee
  Joint Angle Based}} on {{Surface Electromyography Using}} a {{Long Short-Term
  Memory Neural Network}} and {{Time-Advanced Feature}},'' \emph{Sensors},
  vol.~20, no.~17, p. 4966, Sep. 2020.

\bibitem{wangContinuousEstimationHuman2022}
S.~Wang, H.~Tang, L.~Gao, and Q.~Tan, ``Continuous {{Estimation}} of {{Human
  Joint Angles From sEMG Using}} a {{Multi-Feature Temporal Convolutional
  Attention-Based Network}},'' \emph{IEEE Journal of Biomedical and Health
  Informatics}, vol.~26, no.~11, pp. 5461--5472, Nov. 2022.

\bibitem{shiPhysicsInformedLowShotAdversarial2024a}
Y.~Shi, S.~Ma, Y.~Zhao, C.~Shi, and Z.~Zhang, ``A {{Physics-Informed Low-Shot
  Adversarial Learning}} for {{sEMG-Based Estimation}} of {{Muscle Force}} and
  {{Joint Kinematics}},'' \emph{IEEE Journal of Biomedical and Health
  Informatics}, vol.~28, no.~3, pp. 1309--1320, Mar. 2024.

\bibitem{tanejaFeatureEncodedPhysicsInformedParameter2022a}
K.~Taneja, X.~He, Q.~He, X.~Zhao, Y.-A. Lin \emph{et~al.}, ``A
  {{Feature-Encoded Physics-Informed Parameter Identification Neural Network}}
  for {{Musculoskeletal Systems}},'' \emph{Journal of Biomechanical
  Engineering}, vol. 144, no.~12, p. 121006, Dec. 2022.

\bibitem{zhangBoostingPersonalizedMusculoskeletal2023}
J.~Zhang, Y.~Zhao, T.~Bao, Z.~Li, K.~Qian \emph{et~al.}, ``Boosting
  {{Personalized Musculoskeletal Modeling With Physics-Informed Knowledge
  Transfer}},'' \emph{IEEE Transactions on Instrumentation and Measurement},
  vol.~72, pp. 1--11, 2023.

\bibitem{maPhysicsInformedDeepLearning2024a}
S.~Ma, J.~Zhang, C.~Shi, P.~Di, I.~D. Robertson \emph{et~al.},
  ``Physics-{{Informed Deep Learning}} for {{Muscle Force Prediction With
  Unlabeled sEMG Signals}},'' \emph{IEEE Transactions on Neural Systems and
  Rehabilitation Engineering}, vol.~32, pp. 1246--1256, 2024.

\bibitem{zhaoEMGDrivenMusculoskeletalModel2020}
Y.~Zhao, Z.~Zhang, Z.~Li, Z.~Yang, A.~A. {Dehghani-Sanij} \emph{et~al.}, ``An
  {{EMG-Driven Musculoskeletal Model}} for {{Estimating Continuous Wrist
  Motion}},'' \emph{IEEE Transactions on Neural Systems and Rehabilitation
  Engineering}, vol.~28, no.~12, pp. 3113--3120, Dec. 2020.

\bibitem{millard2013flexing}
M.~Millard, T.~Uchida, A.~Seth, and S.~L. Delp, ``Flexing computational muscle:
  modeling and simulation of musculotendon dynamics,'' \emph{Journal of
  biomechanical engineering}, vol. 135, no.~2, p. 021005, 2013.

\bibitem{lloydEMGdrivenMusculoskeletalModel2003}
D.~G. Lloyd and T.~F. Besier, ``An {{EMG-driven}} musculoskeletal model to
  estimate muscle forces and knee joint moments in vivo,'' \emph{Journal of
  Biomechanics}, vol.~36, no.~6, pp. 765--776, Jun. 2003.

\bibitem{thelenAdjustmentMuscleMechanics2003}
D.~G. Thelen, ``Adjustment of {{Muscle Mechanics Model Parameters}} to
  {{Simulate Dynamic Contractions}} in {{Older Adults}},'' \emph{Journal of
  Biomechanical Engineering}, vol. 125, no.~1, pp. 70--77, Feb. 2003.

\bibitem{zajacMuscleTendonProperties}
F.~E. Zajac, ``Muscle and tendon: Properties, models, scaling, and application
  to biomechanics and motor control.''

\bibitem{sethOpenSimSimulatingMusculoskeletal2018}
A.~Seth, J.~L. Hicks, T.~K. Uchida, A.~Habib, C.~L. Dembia \emph{et~al.},
  ``{{OpenSim}}: {{Simulating}} musculoskeletal dynamics and neuromuscular
  control to study human and animal movement,'' \emph{PLOS Computational
  Biology}, vol.~14, no.~7, p. e1006223, Jul. 2018.

\end{thebibliography}


\end{document}